\documentclass[twocolumn,showpacs,prb,superscriptaddress]{revtex4}
\usepackage{graphicx}
\usepackage{amsmath}
\begin{document}

\title{An Analytic Equation of State for Ising-like Models}
\author{Denjoe O'Connor}
\email{denjoe@stp.dias.ie }\affiliation{School of Theoretical Physics,
Dublin Institute for Advanced Studies, 10 Burlington Road, Dublin 4, Ireland}
\author{J.A. Santiago}
\email{santiago@nucleares.unam.mx}
\affiliation{%
Centro de Investigaci\'on Avanzada en Ingenier\'\i a Industrial,
Universidad Aut\'onoma del Estado de Hidalgo,
C.P. 42090 Pachuca, Hgo., M\'exico}
\author{C.R. Stephens}%
\email{stephens@nucleares.unam.mx}
\affiliation{%
Instituto de Ciencias Nucleares,
Universidad Nacional Aut\'onoma de M\'exico,
C.P. 04510 M\'exico, D.F., M\'exico}
\affiliation{School of Theoretical Physics,
Dublin Institute for Advanced Studies, 10 Burlington Road, Dublin 4, Ireland}

\date{\today}

\begin{abstract} 
Using an Environmentally Friendly Renormalization we derive, from an
underlying field theory representation, a formal expression for the
equation of state, $y=f(x)$, that exhibits all desired asymptotic and
analyticity properties in the three limits $x\rightarrow 0$,
$x\rightarrow \infty$ and $x\rightarrow -1$. The only necessary inputs
are the Wilson functions $\gamma_\lambda$, $\gamma_\varphi$ and
$\gamma_{\varphi^2}$, associated with a renormalization of the
transverse vertex functions. These Wilson functions exhibit a
crossover between the Wilson-Fisher fixed point and the fixed point
that controls the coexistence curve. Restricting to the case $N=1$, we
derive a one-loop equation of state for $2< d<4$ naturally
parameterized by a ratio of non-linear scaling fields. For $d=3$ we
show that a non-parameterized analytic form can be deduced. Various
asymptotic amplitudes are calculated directly from the equation of
state in all three asymptotic limits of interest and comparison made
with known results. By positing a scaling form for the equation of
state inspired by the one-loop result, but adjusted to fit the known
values of the critical exponents, we obtain better agreement with
known asymptotic amplitudes.
\end{abstract}


\maketitle

\section{Introduction}

The universal equation of state for the Landau-Ginzburg-Wilson $O(N)$
model remains a subject of great interest (see, for instance,
\cite{zinnjustin} and \cite{Pelisseto} for recent reviews). Its
calculation from first principles is much more difficult than the
calculation of other universal quantities, such as critical exponents
or amplitude ratios. The universal equation of state for the scaling
function exhibits crossover behavior between three distinct,
asymptotic regimes - the critical region when approached from the
critical isotherm, or when approached along the critical isochor, and
the coexistence curve.  Maintaining the correct analyticity properties
of the equation of state in all these three distinct regimes has not
been possible within the confines of first principle calculations,
such as from a field-theoretic, microscopic Hamiltonian. Rather,
appeal has been made to a parameterized phenomenological scaling ansatz
\cite{Schofield} that exhibits the right asymptotics, the underlying
microscopic theory then being used to fix various free parameters that
exist in the ansatz.
 
In this paper we use a renormalization group methodology -
Environmentally Friendly Renormalization \cite{Stephens} - based only
on an underlying Landau-Ginzburg-Wilson Hamiltonian and without the
need for any phenomenological ansatz, to derive an equation of state
that exhibits all desired analyticity properties in the three distinct
asymptotic regimes.

\section{The Equation of State}

As first noted by Widom \cite{Widom}, in the critical region, the
equation of state is a homogeneous function relating an external
magnetic field $H$, the reduced temperature $t$, and the magnetization
$\varphi$, which can be expressed by
\begin{equation}
y=f\left( x\right)  \label{fscaling}
\end{equation}
where $f(x)$ is universal, the scaling variables $y$ and $x$ are
$y=B_c^\delta H/\varphi^{\delta }$ and
$x=B^{1/\beta}t/\varphi^{1/\beta }$, and $B_c$ and $B$ are
non-universal amplitudes associated with the behavior on the critical
isotherm $t=0$ and the coexistence curve $t<0$, $H=0$.

The scaling function $f(x)$ is normalized such that $f(0)=1$ on the
critical isotherm, and $f(-1)=0$ on the coexistence curve.  Several
properties of the universal equation of state are known rigorously.
For instance, it is known that $f(x)$ has a regular Taylor expansion
around the limit $x=0$ given by
\begin{equation}
f(x)=1+\sum_{n=1}^\infty f^0_n x^n ,
\label{expansionzero}
\end{equation} 
while in the limit $x\rightarrow \infty$, by Griffith's analyticity,
one has an expansion of the form
\begin{equation}
\label{expansioninfty}
f(x)=x^\gamma\sum_{n=0}^\infty f^\infty_n x^{-2n\beta}
\end{equation} 
In the limit $x\rightarrow\infty$ a natural variable is
$z=b_1\varphi/t^{\beta}$, where $b_1={(-C_4^+/{(C^+)}^3)}^{1/2}$ and
the $C_{2n}^+$ are the amplitudes of the $2n$-point correlation
functions for $T>T_c$, $C^+$ being the amplitude of the
susceptibility. In terms of $z$ the equation of state takes the form
\begin{equation}
H=-(-C^+/C_4^+)^{1/2}t^{\beta\delta}F(z)
\end{equation}
where the universal scaling function $F(z)$ for small $z$ has an
expansion of the form
\begin{equation}
F(z)=z +\frac{1}{6}z^3+\sum_{n=3}^\infty \frac{r_{2n}}{(2n-1)!}z^{2n-1}
\label{expansioninz}
\end{equation}
where $r_2=r_4=1$ by choice of normalization.  As (\ref{expansioninz})
is an expansion in $\varphi$, the constants $r_{2n}$ are related to
the $2n$-point correlation functions at $\varphi=0$ and hence are very
natural observables to calculate in lattice simulations. In the limit
$z\rightarrow\infty$, $F(z)$ has an expansion of the form
\begin{equation}
F(z)=z^\delta\sum_{k=0}^\infty F_k^\infty z^{-k/\beta}
\end{equation}
The universal scaling functions $f(x)$ and $F(z)$ are related via
\begin{equation}
z^{-\delta}F(z)=F_0^\infty f(x)
\end{equation} 
with $z=z_0x^{-\beta}$, where $z_0$ is the universal zero of the
equation of state in terms of the variable $z$. Hence, the expansion
coefficients of the two functions can be related to find
\begin{eqnarray}
f_n^\infty&=&z_0^{2n+1-\delta}\frac{r_{2n+2}}{F_0^\infty(2n+1)!}\label{fversusr}\\
f_n^0&=&\frac{F_n^\infty}{F_0^\infty}z_0^{-n/\beta}
\end{eqnarray}
Thus, we see it is sufficient to know the expansion coefficients of
$f(x)$ in the limits $x\rightarrow 0$ and $\infty$ in order to
calculate the asymptotic properties of $F(z)$ and the interesting
coefficients $r_{2n}$.

Unlike the limits $x\rightarrow 0$ and $\infty$, near the coexistence
curve, $x\rightarrow -1$, there are no rigorous mathematical arguments
as to the analyticity properties of $f(x)$, although there do exist
conjectures. For instance, for $N>1$ in \cite{Wallace}, based on an
$\varepsilon$-expansion analysis, it was conjectured that $(1+x)$ has
a double expansion in powers of $y$ and $y^{(d-2)/2}$ of the form
\begin{equation}
1+x=c_{1}y+c_{2}y^{1-\epsilon /2}+d_{1}y^{2}+d_{2}y^{2-\epsilon /2}+...
\label{Coexi}
\end{equation}
In three dimensions it predicts an expansion of $(1+x)$ in powers of $y^{1/2}$.

Of course, the behavior in the vicinity of the 
coexistence curve depends on the value of $N$. For $N=1$, the longitudinal 
correlation length remains finite away from the critical point on the 
coexistence curve, while for $N>1$, the existence of Goldstone bosons leads
to infrared singularities. For $N=1$, one may formally posit that in the 
vicinity of the coexistence curve
\begin{equation}
f(x)=\sum_{n=1}^\infty f_n^c(1+x)^n
\label{ansatzcoex}
\end{equation}
the integer powers being a reflection of the finite longitudinal correlation
length. For Ising-like systems essential singularities are to be
expected. Of course, these cannot be captured within the confines of an
ansatz like (\ref{ansatzcoex}). For $N>1$ studies of the non-linear $\sigma$ 
model lead one to expect a leading behavior of the form
\begin{equation}
f(x)\sim c_f(1+x)^{2/(d-2)}
\end{equation}
though, as mentioned, the nature of the corrections to this behavior
is not well understood, although (\ref{Coexi}) is one conjecture. In
the $1/N$ expansion there is some evidence \cite{Pelisseto} for
logarithmic corrections of the form $\ln(1+x)$ in three dimensions.

Early field theoretic calculations using the renormalization group
with an $\varepsilon$-expansion \cite{Brezin} foundered on the fact
that they did not exhibit Griffiths analyticity in the large $x$
limit. Irrespective of the expansion method - $\varepsilon$-,
fixed-dimension, $1/N$ etc. - there will remain a fundamental
difficulty - that an expansion around a particular fixed point will
not readily access the universal equation of state in the entire phase
diagram, due to the presence of other fixed points that must be
accessed. Simply put, the equation of state exhibits crossovers, and
the nature of these crossovers depends on $N$. For $N=1$ the theory is
controlled by a ``Gaussian'' or mean-field fixed point\endnote{Note
however that this regime also corresponds to that of the ``strong
coupling discontinuity'' fixed point \cite{nienhuis,fisher}.}, wherein
fluctuations are suppressed on the coexistence curve away from the
critical point by the non-vanishing longitudinal mass.  In
distinction, for $N>1$, the theory is dominated by the massless
Goldstone excitations on the coexistence curve and the non-linear
$\sigma$ model gives a good description \cite{Lawrie}.

The problem of incorporating Griffiths analyticity was solved by using 
a parameterized formulation in terms of new variables $R$ and $\theta$, related
to $t$ and $\varphi$ via 
\begin{eqnarray}
\varphi=m_0R^{\beta}m(\theta)\nonumber\\
t=R(1-\theta^2)\nonumber\\
H=h_0R^{\beta\delta}h(\theta)\nonumber
\end{eqnarray} 
where the two functions $m(\theta)$ and $h(\theta)$ are undetermined. 
In this new parameterization the scaling variable $x$ and the scaling function
$f(x)$ are given by
\begin{eqnarray}
x &=&\frac{1-\theta ^{2}}{\theta _{0}^{2}-1}\left( \frac{m\left( \theta
_{0}\right) }{m\left( \theta \right) }\right) ^{1/\beta }  \label{scot1} \\
f\left( x\right)  &=&\left( \frac{m\left( \theta \right) }{m\left( 1\right) }%
\right) ^{-\delta }\frac{h\left( \theta \right) }{h\left( 1\right) }
\label{scot2}
\end{eqnarray} 

Most current field theoretic formulations for determining the equation of 
state (see \cite{zinnjustin,Pelisseto} for comprehensive reviews) rely on such
formulations. The drawback is that the underlying microscopic theory is not
used to determine the functional form of $m(\theta)$ and $h(\theta)$, rather
an ansatz is made as to the general functional form which depends on certain
unknown parameters and then the underlying microscopic theory is used to fix
these parameters. The most common ansatz is that the functions are 
polynomials in $\theta$. The coefficients of the powers of $\theta$ are 
then determined by calculating certain observables independently from the
underlying microscopic theory and then using the values of these observables
to determine the coefficients. Various methodologies have been used 
\cite{Pelisseto,Engels,zinnjustin} using a variety of methods. For instance,
an expansion of the effective potential for small $\varphi$ for $t>0$
using $\epsilon$ or fixed dimension expansions. Results from Monte-Carlo  
simulations or high temperature expansions have also been used.

In this paper we apply a new method to describe the equation of state
in a uniform manner. Using environmentally friendly
renormalization\cite{Stephens}, we find a schema able to capture the
crossover between the Wilson-Fisher fixed point and the fixed point
that controls the coexistence curve. By integrating along curves of
constant magnetization, we obtain the equation of state in the whole
critical region. Moreover, the equation of state that we obtain is
parameterized in terms of the inverse of the transverse correlation
length, a quantity well defined over the entire phase diagram. The
representation we have found is valid for both large and small values
of the scaling variables and satisfies Griffiths analyticity. Although
our approach works for any $N$ we here concentrate on the case $N=1$.

\section{A Renormalization Group Representation 
of the Equation of State}

In this section we briefly outline the derivation of the equation of state
for a theory described by the standard LGW Hamiltonian with $O(N)$ symmetry
\begin{eqnarray}
{\cal H}[{\bf\varphi}]=\int\!
d^{d}x\left({1\over 2}\nabla\varphi^a\nabla\varphi^a
\!+{1\over 2}r(x)\varphi^a\varphi^a
\!+{\lambda_B\over 4!}(\varphi^a\varphi^a)^2\!\right).
\label{hamilt} 
\end{eqnarray}
with $r=r_c+t_B$, where $r_c$ is the value of $r$ 
at the critical temperature $T_c$ and $t_B=\Lambda^2\ {(T-T_c)\over T_c}$,
$\Lambda$ being the microscopic scale.
We denote a generic vertex function by $\Gamma^{(N,M)}_{l\dots
lt\dots t}$, where the number of $l$ and $t$ subscripts indicates
whether a longitudinal or a transverse propagator is to be attached to
the vertex at the corresponding point. When all subscripts are either
$l$ or $t$ we will use a single $l$ or $t$, for example
$\Gamma^{(N,M)}_{t\dots t}$ will be abbreviated
$\Gamma^{(N,M)}_t$. Furthermore, when there are no $\varphi^2$
insertions (i.e. $M=0$) the second index will be left off
e.g. $\Gamma^{(N)}$ indicates $\Gamma^{(N,0)}$.

Due to the Ward identities of the model\endnote{This is also true in the 
``analytically continued'' $N=1$ case.}, it is sufficient to know only
the $\Gamma^{(N,M)}_t$, as all the other vertex functions can be
reconstructed from these. The equation of state is 
\begin{eqnarray}
H=\Gamma^{(2)}_t\varphi
\label{basiceqnofstate1}
\end{eqnarray}  

\subsection{Renormalization in terms of Non-Linear Scaling Fields}
\label{nonlin}

Due to the existence of large fluctuations in the critical regime a
renormalization of the microscopic bare parameters of the form
\begin{eqnarray} 
t(m,\kappa)&=&Z_{\varphi^2}^{-1}(\kappa)t_B(m)\label{coord1}\\ 
\lambda(\kappa)&=&Z_\lambda(\kappa)\lambda_B \label{coord2} \\
\varphi(\kappa)&=&Z_{\varphi}^{-1/2}(\kappa)\varphi_B \label{coord3} 
\end{eqnarray}
must be imposed, where $\kappa$ is an arbitrary renormalization scale.
The renormalized parameters satisfy the differential equations
\begin{eqnarray} 
\kappa{dt(\kappa)\over d\kappa}&=&\gamma_{\varphi^2}(\kappa)t(\kappa) 
\label{wilson1} \\
\kappa{d\lambda(\kappa)\over d\kappa}&=&\gamma_\lambda(\kappa)\lambda(\kappa) 
\label{wilson2} \\
\kappa{d\varphi(\kappa)\over d\kappa}&=&-{1\over2}\gamma_\varphi(\kappa)\varphi(\kappa) 
\label{wilson3} 
\end{eqnarray}
where the Wilson functions associated with this coordinate transformation are
\begin{eqnarray}
\gamma_{\varphi^2}(\kappa)&=&-\left.{\kappa{d\over d\kappa}\ln Z_{\varphi^2}}\right\vert_c
\label{gammaphi2_def}\\
\gamma_\lambda(\kappa)&=&\left.{\kappa{d\over d\kappa}\ln Z_{\lambda}}\right\vert_c
\label{gammalambda_def}\\
\gamma_\varphi(\kappa)&=&\left.{\kappa{d\over d\kappa}\ln Z_{\varphi}}\right\vert_c
\label{gammaphi_def}
\end{eqnarray}
and the derivative is taken along an appropriately chosen curve 
in the phase diagram, which we here denote by $c$. Similarly, integration
of the renormalization group equation for any multiplicatively renormalizable 
$\Gamma^{(N,M)}_t$ yields
\begin{eqnarray}
\label{gammanmrge}
\Gamma^{(N,M)}_t(t,\lambda,\varphi)
=\ \ \ \ \ \ \ \ \ \ \ \ \ \ \ \ \ \ \ \ \ \  \ \ \ \ \ \ \ \ \ \ \ \nonumber\\
{\rm e}^{\int^{m_t}_\kappa({N\over2}\gamma_\varphi-M\gamma_{\varphi^2}){dx\over x}}
\Gamma^{(N,M)}_t(t(\kappa),\lambda(\kappa),\varphi(\kappa))
\end{eqnarray}
To impose a specific, as opposed to abstract, coordinate transformation 
between bare and renormalized theory the renormalization constants 
$Z_\varphi$, $Z_{\varphi^2}$ and $Z_\lambda$ must be fixed. Here, we impose
the explicitly magnetization dependent normalization conditions
\begin{eqnarray} 
\left.\partial_{p^2}\Gamma_t^{(2)}(p,t(\kappa,\kappa),\lambda(\kappa),\varphi(\kappa),\kappa)\right\vert_{p^2=0}&=&1 
\label{nctwof} \\
\Gamma_t^{(2,1)}(0,t(\kappa,\kappa),\lambda(\kappa),\varphi(\kappa),\kappa)
&=&1\label{ncthreef} \\
\Gamma_t^{(4)}(0,t(\kappa,\kappa),\lambda(\kappa),\varphi(\kappa),\kappa)
&=&\lambda .\label{ncfourf} 
\end{eqnarray} 
Note that in this case we impose the normalization conditions on the
transverse correlation functions. These conditions serve to fix the three $Z$
functions associated with $\varphi_B$, $t_B$ and $\lambda_B$ while the
condition
\begin{eqnarray} 
\kappa^2=\Gamma_t^{(2)}(0,t(\kappa,\kappa),\lambda(\kappa),\varphi(\kappa),\kappa),
\label{nconef} 
\end{eqnarray} 
serves as a gauge fixing condition that relates the sliding renormalization 
scale $\kappa$ to the physical temperature $t$ and the physical 
magnetization $\varphi$. Physically, $\kappa$ is a fiducial value of the 
non-linear scaling field $m_t$, which is the inverse transverse correlation 
length. 

Besides $m_t$, the other non-linear scaling field we use to 
parameterize our results is 
\begin{equation}
m_{\varphi}^2={1\over3}{\Gamma_t^{(4)}\varphi^2\over 
\partial_{p^2}\Gamma_t^{(2)}\vert_{p^2=0}}
\label{mfdef}
\end{equation}
which is an RG invariant. It represents the anisotropy in the masses 
of the longitudinal and transverse modes and is related to the 
stiffness constant $\rho_s=\varphi^2\partial_{p^2}
\Gamma_t^{(2)}\vert_{p^2=0}$ via $m_{\varphi}^2= {1\over 3}\lambda\rho_s$.

With a given renormalization prescription one may determine the equation 
of state in terms of the non-linear scaling fields $m_t$ and $m_\varphi$,
as the transverse and longitudinal propagators that appear in all perturbative
diagrams can be parameterized in terms of them. One of the main motivations
for this reparametrization in terms of non-linear scaling fields is that
it eliminates all tadpole diagrams at higher loop order.
However, what is required is the equation of state in terms of the linear
scaling fields $t$ and $\varphi$. One must therefore determine the coordinate
transformation, $t=t(m_t,m_\varphi)$, $\varphi=\varphi(m_t,m_\varphi)$, 
between them.

\subsection{Relating Non-linear and Linear Scaling Fields}

This may be done by specifying a particular curve, $c$, in the
phase diagram along which we integrate the differential relation
for the tranverse vertex functions $\Gamma^{(N)}_t$. For example,
\begin{equation}
d\Gamma^{(2)}_t=\Gamma^{(2,1)}_t dt+{1\over6}\Gamma^{(4)}_t d\varphi^2 
\label{diffrelngtwo}
\end{equation}
can be integrated along a curve of constant $\varphi$ to yield $dt= d\Gamma^{(2)}_t/
\Gamma^{(2,1)}_t$, where the right hand side is naturally written in the 
coordinate system $(m_t,m_\varphi)$. To integrate the renormalization group 
equation for the vertex functions it is most natural to use $m_t=\kappa$ as the 
flow variable and hold $m_\varphi$ constant. However, if we then wish to 
integrate (\ref{diffrelngtwo}) along a curve of constant $\varphi$ we must 
include a Jacobian factor, $2/(2-\gamma_\lambda+\gamma_\varphi)$, that takes 
into account the variation of $m_\varphi$ along a constant $\varphi$ curve.  
The relation between $m_\varphi$ and $\varphi$ is specified by (\ref{mfdef})
using the renormalization group equations for $\partial_{p^2}\Gamma_t^{(2)}$
and $\Gamma_t^{(4)}$ with the normalization conditions (\ref{nctwof}) and 
(\ref{ncfourf}). Using (\ref{gammanmrge}) for $\Gamma_t^{(2)}$
and $\Gamma_t^{(2,1)}$ and the normalization conditions (\ref{ncthreef}) and 
(\ref{nconef}) one may write 
\begin{equation}
\frac{d\Gamma^{(2)}_t}{\Gamma^{(2,1)}_t}=(2-\gamma_\varphi)
{\rm e}^{-\int^\kappa_{\kappa_0}\gamma_{\varphi^2}\frac{dx}{x}}\kappa d\kappa
\label{curveint}
\end{equation} 

In the universal limit, where $\lambda\rightarrow\infty$\endnote{This
universal limit can also be accessed by fixing the coupling at its
asymptotic fixed point value.}, the crossover to mean field theory is 
pushed off to infinity and the theory is then controlled in the limit 
$\kappa\rightarrow\infty$ by the Wilson-Fisher fixed point. Hence, 
$\gamma_i\rightarrow \gamma_i^{\rm WF}$, where 
$\gamma_i^{\rm WF}$ is the Wilson function $\gamma_i$ at 
the Wilson-Fisher fixed point. Defining 
$\Delta\gamma_i=(\gamma_i-\gamma_i^{\rm WF})$ (\ref{curveint})
yields
\begin{equation}
dt
=(2-\gamma_\varphi)
{\rm e}^{-\int^\kappa_{\kappa_0}\Delta\gamma_{\varphi^2}\frac{dx}{x}}
\left({\kappa\over\kappa_0}\right)^{-\gamma_{\varphi^2}}\kappa d\kappa
\label{curveinttwo}
\end{equation} 
To integrate (\ref{curveinttwo}) we need to fix some boundary condition.
The coexistence curve, where $m_t=0$, is a natural one. In this case, 
$dt$ integrates to the temperature variable $(T-T_c(\varphi))$, where 
$T_c(\varphi)$ corresponds to that point on the coexistence curve
where the magnetization is $\varphi$. As we wish to use as temperature
variable $t=(T-T_c)$, we may write $(T-T_c(\varphi))= t + \Delta$, where 
$\Delta=(T_c-T_c(\varphi))$ is the temperature shift that measures the 
distance between the critical point and the point on the coexistence
curve, $T_c(\varphi)$. 
In the integral, $-\int^\kappa_{\kappa_0}\Delta\gamma_{\varphi^2}dx/x$,
one may safely take the universal limit, $\kappa_0\rightarrow\infty$.
In this universal scaling limit, in terms of the linear scaling fields 
$t$ and $\varphi$, the problem has only one scaling variable, $x$. In terms
of the coordinates $m_t$ and $m_\varphi$, this manifests itself as a
reduction to the single scaling variable, $z=m_t/m_\varphi$. Passing to 
this variable, using (\ref{mfdef}), 
and integrating this along a curve of constant $\varphi$ (taking into 
account the Jacobian factor) one finds  
\begin{eqnarray}
A_1(1+x)&=&\mathcal F(z)
\label{fundrelone}
\end{eqnarray}
where the scaling variable $x=B^{1/\beta}t/\varphi^{1/\beta}$, $B$ being the 
non-universal amplitude introduced previously, and the universal scaling 
function $\mathcal F(z)$ is
\begin{eqnarray}
\mathcal F(z)=
\int_{0}^{z}\frac{2(2-\gamma _{\varphi })}{2-\gamma _{\lambda }
+\gamma _{\varphi }}%
D(x)
x^{1\over\beta }\frac{dx}{x} \nonumber \label{te} 
\end{eqnarray}
where
\begin{equation}
D(x)=\exp\left({-\int_{\infty }^{x}2
\left({\frac{\Delta \gamma _{\varphi ^{2}}-\frac{\Delta \gamma _{\lambda }%
}{2\beta }+\frac{\Delta \gamma _{\varphi }}{2\beta }}
{2-\gamma _{\lambda }+\gamma _{\varphi }%
}}\right) \frac{dy}{y}%
}\right)
\end{equation}
The quantity $A_1$ is related to the universal zero of the equation of state.
In terms of the amplitude $B$ 
\begin{equation}
B^2=\frac{\lambda}{3\kappa^{(4-d-\eta)}}A_1^{2\beta}
\label{bandaone}
\end{equation} 

Equation (\ref{fundrelone}) determines the coordinate transformation,
$t=t(m_t,m_\varphi)$ $\varphi=\varphi(m_t,m_\varphi)$, in the scaling
limit where there is only one relevant scaling variable associated
with the linear scaling fields, $x\sim t/\varphi^{1/\beta}$, and one
relevant scaling variable associated  with the non-linear scaling
fields, $z=m_t/m_\varphi$. Hence, we determine the coordinate
transformation $x=x(z)$. Note that the geometry of this transformation 
has some unusual properties relative to the $(T,\varphi)$ plane. The
coexistence curve is mapped to the single point $x=-1$, the critical
isotherm is mapped to the single point $x=0$, and the critical 
isochor, for $t>0$, maps to the single point at infinity. However, the
critical point itself - as the intersection of the coexistence curve,
critical isotherm and critical isochor - maps to all these points.

To proceed to the equation of state we use (\ref{basiceqnofstate1}) 
and the renormalization group equation for $\Gamma_t^{(2)}$. Once 
again, passing to the variable $z$ and including in the Jacobian 
factor, one obtains
\begin{eqnarray}
{H\over \varphi^\delta}=
\left(\frac{\lambda}{3\kappa^2}\right)^{(\delta-1)\over 2}{\mathcal G(z)}
\end{eqnarray}
where the universal scaling function $\mathcal G(z)$ is given by
\begin{eqnarray}
\mathcal G(z) =z^{\frac{\gamma }{\beta }}{\rm e}^{\frac{\gamma }{%
\beta }\int_{\infty }^{z}\frac{\Delta \gamma _{\lambda }-\Delta \gamma
_{\varphi }}{2-\gamma _{\lambda }+\gamma _{\varphi }}\frac{dy}{y}%
}{\rm e}^{-\int_{\infty }^{z}\frac{2\Delta \gamma _{\varphi }}
{2-\gamma _{\lambda}+\gamma _{\varphi }}\frac{dy}{y}}  
\label{ache}
\end{eqnarray}
Introducing the scaling variable $y=A_3^{-1}(H/\varphi^\delta)$, where $A_3$ 
is related to the non-universal amplitude $B_c$ via
\begin{equation}
B_c^\delta=A_3\left(\frac{\lambda}{3\kappa^2}\right)^{(\delta-1)\over 2}
\label{bcandathree}
\end{equation}
one sees that the universal equation of state is now of the form 
(\ref{fscaling}) with
\begin{equation}
f(x)=\frac{1}{A_3}\mathcal G({\mathcal F}^{-1}(A_1(1+x))
\end{equation}

We may now ask what more can be said about $A_1$ and $A_3$, or whether they
are simply non-universal parameters, being related to the non-universal
amplitudes $B_c$ and $B$, that cannot be determined within the present formalism?
$A_1$, in fact, is related to the universal zero of the equation of state, and 
may be calculated in the following manner: We choose some arbitrary value
of $z$, $z_0$, and write $\mathcal F(z) \rightarrow \mathcal F(z_0)
+\int_{z_0}^z F(x)dx$; then choose $z$ and $z_0$ in the asymptotic 
regime $z$, $z_0\rightarrow\infty\ \ z>z_0$, wherein, from Griffith's 
analyticity, we can write $\mathcal F(z)= z^{1/\beta}\sum_{n=0}^\infty
\mathcal F^\infty_n z^{-2n}$. In the limit $z\rightarrow \infty$, depending
on the value of $\beta$, certain terms in the asymptotic expansion diverge.
For instance, for the three dimensional Ising model, at one loop, $\beta=3/10$,
hence, only the $n=0$ and $n=1$ terms diverge, while the contributions from
$n\geq 2\rightarrow 0$. Denoting the divergent part of the expansion as 
$\mathcal F^\infty(z)$; as $z_0$ is constant we may identify the non-constant
term, $A_1x$, in this limit with $\mathcal F^\infty(z)$. Hence, we may 
identify $A_1=\lim_{z_0\rightarrow\infty}(\mathcal F(z_0)-
\mathcal F^\infty(z_0))$. In terms of the integrand, $I(x)$, of the scaling
function $\mathcal F$, we may write
\begin{equation}
A_1=\int_0^\infty(I(x)-I^\infty(x))dx
\label{aone}
\end{equation} 
where $I^\infty(x)$ is defined via $\mathcal F^\infty=\int_0^zI^\infty(x)dx$.
$A_1$ is clearly universal. To determine $A_3$, we set the condition 
$y =1$ on the critical isotherm $t=0$. This corresponds to
a particular value, $z_c$, of $z$. Hence, $A_3=\mathcal G(z_c)$. To determine 
$z_c$ note that from (\ref{te}) on the critical isotherm 
$A_1=\mathcal F(z_c)$. The inversion of this function allows for the 
identification of $z_c$. Hence, we deduce $A_3$ to be
\begin{equation}
A_3=\mathcal G(\mathcal F^{-1}(A_1))
\label{athreefromaone}
\end{equation} 
which is, once again, a universal function. 

So, given that $A_3$ is determined from $A_1$, and $A_1$ is determined 
from $\mathcal F$ by subtracting off its divergent component as 
$z\rightarrow\infty$, we see that the simple
ingredients that enter into a complete specification of the universal 
equation of state are the three Wilson functions - $\gamma_{\lambda}$,
$\gamma_{\varphi^{2}}$ and $\gamma_{\varphi}$. These are the only 
quantities that need to be determined perturbatively (or otherwise).

Note that this equation of state has been determined from a first principles
calculation based on an underlying microscopic model. It is parameterized,
but parameterized in a way that is completely determined by the underlying 
model. This is in distinction to standard parametric representations \cite{zinnjustin,Pelisseto},
where, after imposing certain analyticity requirements, there is a large 
arbitrariness in determining the scaling functions $h(\theta)$ and $m(\theta)$.
In fact, these functions may depend on an arbitrary number of parameters and,
for each parameter, a universal quantity must be independently calculated
in order to fix it. Also, the parameter $z$ here has a much more transparent and
direct physical meaning than $\theta$, being simply the ratio of the two 
fundamental non-linear scaling fields in the problem - $m_t$ and $m_\varphi$ - 
the transverse correlation length and the stiffness constant. As these
quantities are well defined throughout the phase diagram this formulation
has an added advantage relative to parametric, fixed-dimension expansions
where the relevant non-linear scaling field used is the mass for $T>T_c$
and there are difficulties reaching the ordered phase \cite{zinnjustin}.

In order to determine the expansion coefficients $f_n^0$ and $f_n^\infty$,
as introduced in section 1, one requires the Taylor expansion of $f(x)$ around 
$x=0$ and $x=\infty$. In terms 
of our parametric representation, $d^nf(x)/dx^n$ can be expressed using 
$d/dx=(dz/dx)d/dz$, where $dz/dx=A_1/({d\mathcal F(z)/dz})$ hence,
\begin{equation}
{d^nf(x)\over dx^n}={A_1\over A_3}
\left(\left({d\mathcal F(z)\over dz}\right)^{-1}{d\over dz}\right)^n\mathcal G(z)
\end{equation}
which need to be evaluated at the points of interest $z=z_c$ ($x=0$), 
$z=\infty$ ($x=\infty$) and $z=0$ ($x=-1$). For instance, taking the limit 
$x\rightarrow\infty$ in (\ref{te}) and (\ref{ache}), and using the fact that
the Wilson functions $\gamma_i$ approach their values at the Wilson-Fisher fixed
point so $\Delta\gamma_i\rightarrow 0$, one finds
\begin{eqnarray}
\mathcal F(z) \rightarrow \gamma z^{1/\beta}\\
\mathcal G(z) \rightarrow z^{\gamma\over\beta}
\end{eqnarray}
Hence,
\begin{equation}
f(x)\rightarrow \frac{A_1^\gamma}{A_3}\gamma^{-\gamma}x^\gamma + O(x^{\gamma-2\beta})
\end{equation}
from which, using (\ref{athreefromaone}) we may identify the expansion coefficient
\begin{equation}
f_0^\infty=\frac{(A_1/\gamma)^\gamma}{\mathcal G(\mathcal F^{-1}(A_1))}
\label{fzeroinf}
\end{equation}
which is related to the universal amplitude ratio $R_\chi$ via 
$f_0^\infty=R_\chi^{-1}$. Using the expression for $C^+$ in our formulation,
$C^+=\kappa^{-(2-\eta)}\gamma^{-\gamma}$ and equations (\ref{bandaone}) and
(\ref{bcandathree}), one can verify that (\ref{fzeroinf})
is identical to the expression $R_\chi=(C^+/B_c)(B/B_c)^{\delta-1}$.
Similarly, using the expression for $C_4^+$ in our formulation
\begin{equation}
C_4^+=-\lambda\gamma^{2\gamma+d\nu}\kappa^{4\gamma-8+2d\nu}
\end{equation}
one may determine the universal amplitude ratio $R_4^+=-C_4^+B^2/(C^+)^3$
to be
\begin{equation}
R_4^+=\frac{3(\mathcal F(z_c))^{2\beta}}{\gamma^{\gamma-d\nu}}
\end{equation}
With these two amplitude ratios in hand two-scale factor universality
implies that any other may be determined.

\section{One-loop Results}

The advantages of the present formulation can be best illustrated by considering
a concrete example. We will consider the universal equation of state in the one 
loop approximation, as at this level it is still possible to obtain analytic or 
``quasi''-analytic results.

We begin with the values of the Wilson functions to one loop. The running 
dimensionless coupling $\lambda$ satisfies
\begin{equation}
z\frac{d\lambda(z)}{dz}=-\varepsilon\lambda +c_d\lambda^2(z)
\left((1+{1\over z^2})^{d-6\over2}+{(N-1)\over9}\right)
\label{betafunction}
\end{equation}
where $c_d=3(4-d)\Gamma((4-d)/2)/2(4\pi)^{d/2}$. Taking the initial condition 
$\lambda(z_0)=\lambda$, in the limit $z_0\rightarrow\infty$, 
$\lambda\rightarrow\infty$ one arrives at the universal separatrix 
solution\endnote{This solution may also be reached by choosing the initial
coupling to be on the separatrix solution at $z=z_0$.}  
\begin{equation}
\lambda(z)=\left(c_d\left((1+{1\over z^2})^{d-6\over2}
+{(N-1)\over9}\right)\right)^{-1}
\end{equation}
On the separatrix
\begin{eqnarray}
\label{wilsononeloopn}
\gamma_\lambda&=&(4-d)\left({(1+{1\over z^2})^{d-6\over2}+{(N-1)\over9}\over
(1+{1\over z^2})^{d-4\over2}+{(N-1)\over9}}\right)\\
 \gamma _{\varphi ^{2}}&=&
(4-d)\left({(1+{1\over z^2})^{d-6\over2}+{(N-1)\over3}\over
3(1+{1\over z^2})^{d-4\over2}+{(N-1)\over3}}\right)\\
\gamma_\varphi&=&0
\end{eqnarray}
In the limit $z\rightarrow\infty$, the Wilson-Fisher fixed point is 
approached and $\gamma_i\rightarrow\gamma^{\rm WF}_i$ with, at one loop,
$\gamma_\lambda=(4-d)$ and $\gamma _{\varphi ^{2}}=(4-d)(N+2)/(N+8)$.
On the contrary, in the limit $z\rightarrow 0$ the strong-coupling 
fixed point is approached and $\gamma_i\rightarrow\gamma^{\rm SC}_i$.
For $N>1$ the Goldstone bosons dominate and 
$\gamma_\lambda=\gamma _{\varphi ^{2}}=(4-d)$. For $N=1$ however, 
this fixed point is mean-field like as fluctuations are suppressed and
$\gamma_i\rightarrow 0$. 

In the limit $z\rightarrow\infty$, the Wilson 
functions can be expanded as power series in $z^{-2}$ for any $N$
\begin{equation}
\gamma_i(z)=\gamma^{WF}_i+\sum_{n=1}^\infty a_i(n)z^{-2n}
\end{equation}
Hence, the universal scaling functions $\mathcal F$ and $\mathcal G$
can also be written as power series in $z^{-2}$. This is true in a 
diagrammatic expansion to all orders not just at one loop. The limit 
$z\rightarrow 0$ is more complicated. In this limit, 
$\gamma_i\rightarrow\gamma_i^{SC}$ but the nature of the corrections
is not obvious. At the one-loop level, from (\ref{wilsononeloopn}), one
can see that the leading corrections to the strong coupling
fixed point values will be $z^{(4-d)/2}$.

For $N\neq 1$ analytic progress is difficult. However, for $N=1$ these
expressions simplify greatly yielding
\begin{eqnarray}
\label{wilsononeloopnone}
\gamma_\lambda=(4-d)\left(1+{1\over z^2}\right)^{-1}\\
\gamma _{\varphi ^{2}}=\frac{(4-d)}{3}\left(1+{1\over z^2}\right)^{-1}
\end{eqnarray}

With these expressions one can explicitly calculate the scaling functions 
$\mathcal G(z)$ and $\mathcal F(z)$ 
\begin{eqnarray}
\label{Gonelooparbitraryd}
\mathcal G(z)& =& z^{2}
\left({2\over(d-2)}+z^2\right)^{(4-d)\over(d-2)}\\
\mathcal F(z)&=&{ 3\over d+2 }\Bigg( \left( {2\over d-2} + z^2 \right)^{2(4-d)\over 3(d-2)}(2z^2-1)\nonumber\\
&+&\left( 2\over d-2\right)^{2(4-d)\over 3(d-2)}\Bigg) 
\label{Fonelooparbitraryd}
\end{eqnarray}
To determine the constant $A_1$ we take the $z\rightarrow\infty$ limit of 
(\ref{Fonelooparbitraryd}), identify the divergent part with $A_1x$ and the
constant remainder with $A_1$.\endnote{If $2(4-d)/3(d-2)$ is a positive integer, 
$n$, then this remainder is zero and $A_1$ cannot be determined by looking at the 
asymptotic limit $z\rightarrow\infty$. This is a pure artefact of the
one-loop approximation, where $\eta=0$, and has no physical meaning.} 
Note that the case $d=2$ is problematic. This is an artefact of 
the one-loop approximation where, in particular, in $d=2$ one has $\beta=0$, which
is clearly unphysical. The case $d=4$ is also inaccessible due to the fact that
we took the universal limit $\lambda\rightarrow\infty$. It may be recovered by
returning to (\ref{betafunction}) and integrating it and not taking this limit.
In this case, as expected, logarithms appear. So, equations (\ref{Gonelooparbitraryd})
and (\ref{Fonelooparbitraryd}) are valid for $2<d<4$. 
$A_3$ can be determined as a function of the undetermined $A_1$ using 
(\ref{athreefromaone}). Explicitly, taking the $z\rightarrow\infty$ limit 
$A_1$ can be identified to be
\begin{equation}
A_1={3\over d+2}\left( 2\over d-2 \right)^{2(4-d)\over 3(d-2)}
\end{equation} 
from which one determines $z_c=2^{-1/2}$ which, interestingly, is dimension
independent, though, once again, this is a one-loop artefact. With $z_c$ in hand
one determines $A_3$ to be
\begin{equation}
A_3=\frac{1}{2}\left({d+2\over 2(d-2)}\right)^{(4-d)\over (d-2)}
\end{equation} 
Thus, the one-loop equation of state for $N=1$ in $d$-dimensions is
\begin{eqnarray}
y &=&{2^{2\over(d-2)}\over 
(d+2)^{(4-d)\over(d-2)}}z^{2}
(2+(d-2)z^2)^{(4-d)\over(d-2)}\label{threedone}\\
x&=&(2z^2-1)(1+(d-2)z^2/2)^{2(4-d)\over 3(d-2)}\label{threedtwo}
\end{eqnarray}
Griffiths analyticity can be seen quite simply from these expressions. For 
$x\rightarrow \infty$ one has $z\rightarrow \infty$ and the right hand side of 
(\ref{threedtwo}) takes the form $z^{1/\beta}\sum_{n=0}^\infty (c_n/z^{2n})$, where
$\beta=3(d-2)/2(d+2)$. Substituting into (\ref{threedone}), one then has the expansion
(\ref{expansioninfty}), where $\gamma=6/(d+2)$. Similarly, by expanding (\ref{threedtwo}) 
around $z=z_c$ one obtains the expansion (\ref{expansionzero}). Finally, in the 
vicinity of the coexistence curve, from (\ref{threedtwo}) we see that $(1+x)$ can 
be written as a power series expansion in $z^2$ and, hence, we verify (\ref{ansatzcoex}).
Once again, within the approximation used one would not expect essential singularities
to appear. To summarize: (\ref{threedone}) and (\ref{threedtwo}) have been determined 
from first principles from the underlying microscopic theory and obey all analyticity
and other properties required in the different asymptotic regimes. 

>From these expressions one may calculate analytically the coefficients
$f_n^0(d)$ and $f_n^\infty(d)$ as functions of $d$, as well as derived quantities, 
such as the $r_{2n}$, and universal amplitude ratios, like $R_\chi$ and $R_4^+$.
In the same way, we may examine the behavior in the vicinity of the coexistence curve. 
First, we determine $z$ as a power series in $(1+x)$, and 
then we replace the result in the expansion of $y$ for $z$ small. 
 
In Table \ref{U1} we show the coefficients $f^0_1(d)-f^0_{5}(d)$ and in Table \ref{U2}
$f^c_1(d)-f^c_3(d)$. The coeffients 
$f^\infty_0(d)-f^\infty_{5}(d)$ for arbitrary $d$ appear in Table \ref{U3}.
Our results for some important quantities like $r_{2n}(d)$  appear in Table \ref{U4}.  
Obviously with analytic expressions in hand it is straightforward to generate other
coefficients.

\begin{table}[h]
\caption{\label{U1}\small Values of $f_n^0$ for the d-dimensional Ising universality class.}
\begin{ruledtabular}
\begin{tabular}{l|r}
 $f_1^0$ & $3\times\,2^{-\frac{1}{3} + 
\frac{8}{3\,\left( d-2 \right) }}\,{\left( 2 + d \right) }^{\frac{2 + d}{6 - 3\,d}}$\\ 

\hline\hline           
$f_2^0$ & $-\left( d-4 \right) \,
\left( 2 + d \right)^{\frac{4 + 2\,d}{6 - 3\,d}}
2^{\frac{8\,\left( 4-d \right) }{3\,\left( d-2 \right) }}$\\

\hline\hline
$f_3^0$ & $(1/3)\times\, 2^{-3 + \frac{8}{d-2}}\,\left( d-4 \right) \,
{\left( 2 + d \right) }^{\frac{2 + d}{2 - d}}$\\

\hline\hline
$f_4^0$ & $(1/27)\times\, 2^{\frac{70 - 19\,d}{3\,d-6}}\,\left( d-7 \right) \,
\left( d-4 \right) \,
{\left( 2 + d \right) }^{\frac{8 + 4\,d}{6 - 3\,d}}\,
\left( 8 + d \right) $\\

\hline\hline
$f_5^0$ & $\frac{\left( d-16 \right) \,\left( d-4 \right) \,
\left( 5 + d \right) \,\left( 2\,d-11 \right) }
{405\times\,
2^{\frac{20\,\left( d-4 \right) }
{3\,\left( d-2 \right) }}\,
{\left( 2 + d \right) }^
{\frac{5\,\left( 2 + d \right) }{3\,\left( d-2 \right) }}}$\\

\end{tabular}
\end{ruledtabular}
\end{table}

\begin{table}[h]
\caption{\label{U2}\small Values of $f_n^c$ for the d-dimensional Ising universality class.}
\begin{ruledtabular}
\begin{tabular}{l|l}

$f_1^{\rm c}$ &  $3\cdot 2^{\frac{6 - d}{d-2}}
\left( 2 + d \right)^{\frac{2}{2-d}}$\\

\hline\hline

$f_2^{\rm c}$ & $3\cdot 2^{\frac{10 - 4d}{d-2}}
    \left( -3\cdot 2^{\frac{2d-2}{d-2}} + 
      5\cdot 2^{\frac{d}{d-2}} \right) {d-4\over (2+d)^2}\left( 2+d\right)^{4-d\over d-2}$\\

\hline\hline

$f_3^{\rm c} $ &  ${64}^{\frac{d-1}{2 - d}} {4-d\over (2+d)^3} 
       {\left( 2 + d \right) }^{\frac{4-d}{d-2}}\,
    \Bigg( 81\cdot 2^{\frac{3d+4}{d-2}} + 
      47\cdot 2^{\frac{2 + 4d}{d-2}} -$\\ 
      & $45\cdot {32}^{\frac{d}{d-2}} -
        35\cdot 2^{\frac{2\,\left( 3 + d \right) }{d-2}}\,d + 
      9\cdot 2^{\frac{2 + 4\,d}{d-2}}\,d \Bigg) $\\

\end{tabular}
\end{ruledtabular}
\end{table}

One may also recover the well known $\varepsilon$-expansion result by substituting 
$d=4-\varepsilon$ in (\ref{threedone}) and (\ref{threedtwo}), obtaining 
\begin{eqnarray}
y&=&{2^{2\over(2-\varepsilon)}\over 
(6-\varepsilon)^{\varepsilon\over(2-\varepsilon)}}z^{2}
(2(1+z^2)-\varepsilon z^2)^{\varepsilon\over(2-\varepsilon)}\label{threedoneepsilon}\\
x&=&(2z^2-1)(1+(2-\varepsilon)z^2/2)^{2\varepsilon\over 3(2-\varepsilon)}
\label{threedtwoepsilon}
\end{eqnarray}
Expanding in powers of $\varepsilon$ to $O(\varepsilon)$ one finds
\begin{eqnarray}
y&=&2z^2\left(1+\frac{\varepsilon}{2}
\ln {2(1+z^2)\over 3}\right)
\label{threedoneexp}\\
1+x&=&2z^2+\frac{\varepsilon}{3}(2z^2-1)\ln (1+z^2)
\label{threedtwoexp}
\end{eqnarray}
Inverting (\ref{threedtwoexp}) in powers of $\varepsilon$ and substituting into 
(\ref{threedoneexp}) one finds
\begin{equation}
y=1+x+\frac{\varepsilon}{2}(1+x)\ln \frac{(x+3)}{3}
-\frac{\varepsilon}{3}x\ln \frac{(x+3)}{2}
\label{epsilonexpresult}
\end{equation}
which is the well known result in terms of normalized variables.

\subsection{The Case $d=3$}

For $d=3$ things are even more transparent, as the parameter $z$ can be totally
eliminated, thereby ending with a direct, unparameterized relation between $x$ 
and $H/\varphi^\delta$. For $d=3$ the scaling functions $\mathcal F(z)$ and
$\mathcal G(z)$ are

\begin{eqnarray}
\label{foneloopthreed}
\mathcal F(z)&=&
\frac{2^{2\over3}3}{5}+\frac{3}{5}(z^2+2)^{2\over3}(2z^2-1)\\
\mathcal G(z)&=&(z^4+2z^2)
\end{eqnarray}
while $A_1$ and $A_3$ are given by  
\begin{equation}
\label{aoneoneloopthreed}
A_1=\frac{2^{2\over3}3}{5}\ \ \ \ \ \ \ \ \ A_3=\frac{5}{4}
\end{equation}

With these values for the amplitudes $A_1$ and $A_3$ the equation of state
in terms of the variable $z$ is 
\begin{equation}
z^4+2z^2-\frac{5y}{4}=0
\end{equation}
which can be simply solved (the positive square root is required) and
substituted into (\ref{foneloopthreed}) to find
\begin{equation}
\label{univeqnofstatethreed}
2^{2\over3}x=
\left(1+\left(1+\frac{5y}{4}\right)^{1\over2}\right)^{2\over3}
\left(2\left(1+\frac{5y}{4}\right)^{1\over2}-3\right)
\end{equation}
which can be seen to satisfy $y=1$ at $x=0$ and $y=0$ at
$x=-1$. For large $x$, $y\rightarrow (2^{8/5}/5)x^{6/5}$, i.e. $y\sim x^\gamma$
as required by Griffith's analyticity. It is valid for both $t>0$ and $t<0$.  
\begin{table}[h]
\caption{\label{U3}\small Values of $f_n^\infty$ for the d-dimensional Ising universality class.}
\begin{ruledtabular}
\begin{tabular}{l|l}

$f_0^\infty$ &  $ \frac{4^{\frac{1}{d-2}}\,{\left( \frac{2 + d}{d-2} \right) }^{\frac{d-4}
{d-2}}}{{\left( 2^{\frac{14 - 5\,d}{6 - 3\,d}}\,
{\left( d-2 \right) }^{\frac{8 - 2\,d}{3\,d-6}} \right) }^{\frac{6}{2 + d}}}  $\\

\hline\hline

$f_1^\infty$ & $ {4^{\frac{1}{d-2}}\over d-2}\,{\left( 2^{\frac{14 - 5\,d}{6 - 3\,d}}\,
{\left( d-2 \right) }^{\frac{8 - 2\,d}{3\,d-6}} \right) }^
{\frac{3\,\left( d-4 \right) }{2 + d}}\,
{\left( \frac{2 + d}{d-2} \right) }^{\frac{d-4}{d-2}} $\\

\hline\hline

$f_2^\infty $ & $ -3\times 2^{\frac{6 - 2d}{d-2}}\left( d-4\over d-2 \right) 
{\left( 2^{\frac{14 - 5\,d}{6 - 3d}}
{\left( d-2 \right) }^{\frac{8 - 2d}{3d-6}} \right) }^
{\frac{6\left( d-3 \right) }{2 + d}}
{\left( \frac{2 + d}{d-2} \right) }^{\frac{d-4}{d-2}}  $\\

\hline\hline

$f_3^\infty$ & $2^{\frac{2(3-d)}{d-2}}{ (d-4)(4d-13)\over (d-2)^3}  
{\left( 2^{\frac{14 - 5\,d}{6 - 3d}}
{\left( d-2 \right) }^{\frac{8 - 2d}{ 3d-6}} \right) }^{\frac{9d-24}{2 + d}}
{\left( \frac{2 + d}{d-2} \right) }^{\frac{d-4}{d-2}} $\\

\hline\hline

$f_4^\infty$ & $2^{\frac{12-5d}{d-2}}{d-4\over (d-2)^4}  
{\left( 2^{\frac{14 - 5d}{6 - 3\,d}}
{\left( d-2 \right) }^{\frac{8 - 2d}{3d-6}} \right) }^{\frac{12d-30}{2 + d}}
{\left( \frac{2 + d}{d-2} \right) }^{\frac{d-4}{d-2}}\times$\\ 
& $\left(17- 5d \right) \left( 11d -32\right)  $\\

\hline\hline

$f_5^\infty$ & ${2^{\frac{2(8d-35)}{d-2}}(d-2)^{\frac{50 - 15d}{d-2}}
{\left( \frac{2 + d}{d-2} \right) }^{\frac{d-4}{d-2}}\over 
-45\left( 2^{\frac{14 - 5d}{6 - 3d}}
{\left( d-2 \right) }^{\frac{8 - 2d}{3d-6}} \right)^{66\over 2+d}}\times\Bigg( 
176553\times 2^{\frac{6d+8}{d-2}}\,d\, + $\\
& $27\left( 2419d-11973 \right)d^2 2^{\frac{4\left( 3 + d \right) }{d-2}}- 
189\left( 1520 + 13d^4 \right)2^{\frac{5\left(2 + d \right) }{d-2}}\Bigg)$  \\
\end{tabular}
\end{ruledtabular}
\end{table}

\begin{table}[h]
\caption{\label{U4}\small Values of $R_4^+, F_0^\infty$ and some $r_{2n}$ coefficients 
for the d-dimensional Ising universality class.}
\begin{ruledtabular}
\begin{tabular}{l|l}

$R_4^+$ & ${6\over d-2}{\left( 2^{\frac{5}{3} + \frac{4}{6 - 3\,d}}\,
{\left( d-2 \right) }^{\frac{8 - 2\,d}{3\,d-6}}
\right) }^{\frac{3\,\left( d-2 \right) }{2 + d}}$\\

\hline\hline           
$F_0^\infty$ & $\frac{4^{\frac{d-1}{2 - d}}\,{\left( d-2 \right) }^2\,
{\left( \frac{2 + d}{d-2} \right) }^{\frac{4 - d}{d-2}}}
{9\,{\left( 2^{\frac{14 - 5\,d}{6 - 3\,d}}\,
{\left( d-2 \right) }^{\frac{8 - 2\,d}{3\,d-6}}
\right) }^{\frac{6\,\left( d-3 \right) }{2 + d}}}$\\

\hline\hline

$r_6$ & ${5\over 2}\left(4-d \right) \,\left( d-2 \right) $\\

\hline\hline

$r_8$ & ${35\over 6}\left(d-4 \right) \,\left(  4\,d-13 \right) $\\

\hline\hline

$r_{10}$& ${35\over 4}\left( 4-d \right) \,\left( 5\,d -17\right) \,
\left( 11\,d -32\right)$\\

\hline\hline

$r_{12} $ & $ 385\times 2^{\frac{2 - 6\,d}{d-2}}\,\Big( 665\times 2^{4 + \frac{5\,d}{d-2}} - 6539\times  2^{6 + 
\frac{10}{d-2}}\,d +$\\ 
& $11973\times 2^{\frac{2 + 4\,d}{d-2}}\,d^2 
- 2419\times 2^{\frac{2 + 4\,d}{d-2}}\,d^3 + 91\times {32}^{\frac{d}{d-2}}\,d^4 \Big)$

\end{tabular}
\end{ruledtabular}
\end{table}

In the second column of Table \ref{U5} we see the numerical values of
the coefficients $f_n^0$ and $f_n^\infty$, as well as some important
derived quantities, such as the $r_{2n}$.  All these values are as in
good an agreement with known values as one might expect from a
one-loop calculation, in those cases where a comparison can be made,
with one apparent exception: the value of $r_8$ is about $2-3$ times
bigger than the majority of estimates, which are in the range
$2.18-2.7$.  However, a Monte Carlo simulation of Kim and Landau
\cite{kimlandau} led to a value of $r_8$ which was much larger than
other estimates. However, this estimate goes down when we fit with the
exact exponents in the next section. Our value of $r_{10} = -17.5$ -
is in the expected region of previous calculations, which give
estimates in the region $-4--25$, though the errors associated with
many of these estimates are large. Our estimate of $r_{12}$ for $d=3$
would seem to be new.  Obviously an important advantage of the present
approach is the facility with which the $r_{2n}$ can be calculated for
even very high $n$.  In the figure we see a comparison between our
one-loop equation of state and that obtained by the HT method.


\begin{figure}[h]
\begin{center}
\includegraphics[width=3.5in,angle=0]{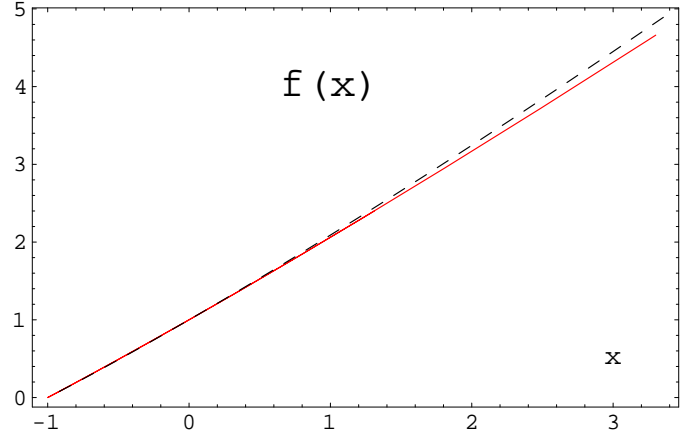}
\end{center}
\caption{\small The  scaling function $f(x)$ of the
three-dimensional Ising model. The dashed line was taken from reference \cite{Guida}.
}   
\label{fscalinggraph}
\end{figure}

\section{Fitted Exponents}
\label{fitted}

Although one of the chief advantages of the present methodology is the fact
that it is completely self-contained, in that there are no parameters to be fixed
by appealing to exogenous information, as in standard parametric approaches,
it is possible adapt the present method to utilize information that is available,
such as precise estimates for the critical exponents and amplitude ratios. To
illustrate this here, we once again consider the case $N=1$. We take the 
one-loop ``crossover'' function $(1+z^{-2})$ to give the exact form of the crossover 
and fit its asymptotic value to the best estimates for the critical exponents. Hence,
we take
\begin{equation}
\gamma_\lambda=\frac{(4-d)}{(1+{1\over z^{2}})} \ \ \ \ 
\gamma_{\varphi^2}=\frac{(2-1/\nu)}{(1+{1\over z^{2}})}\ \ \ \ 
\gamma_{\varphi}=\frac{\eta}{(1+{1\over z^{2}})}
\label{fittedwilson}
\end{equation}
The crossover form for $\eta$ is of course a pure supposition given that the form
$(1+z^{-2})$ is derived from a one-loop calculation. The form of (\ref{fittedwilson})
is such as to guarantee a crossover to the known asymptotic behavior. The same
procedure could be carried out using a two-loop calculation. In this case the 
crossover functions for each Wilson function would be different, due to the fact that
different diagrams contribute to them. Once again constants would be 
introduced to ensure a crossover in the limit $z\rightarrow\infty$ to the correct
exponent values for $\nu$ and $\eta$. With the ansatz (\ref{fittedwilson}) one finds
\begin{eqnarray}
&&\mathcal F(z) = \left(\frac{\nu}{\beta}\right)^{1\over 2\beta}{2\beta(\gamma-1)\over
1-2\beta}\nonumber\\
&&\left[\left({\gamma(1-2\beta)\over 2\nu(\gamma-1)}z^2-1\right)
\left(1+{\beta\over\nu}z^2\right)^{{1\over 2\beta}-1}+1\right]
\label{ffitted}\\
&& \mathcal G(z) = z^{\gamma\over\beta}
\left(1+{\nu\over\beta z^2}\right)^{(\gamma-2\beta)\over 2\beta}
\label{gfitted}
\end{eqnarray}
Examining the large $z$ limit one determines the universal amplitude $A_1$ to be
\begin{equation}
A_1=\left(\frac{\nu}{\beta}\right)^{1\over 2\beta}{2\beta(\gamma-1)\over
1-2\beta}
\label{aonefitted}
\end{equation}
Consequently, we determine $z_c^2=2\nu(\gamma-1)/\gamma(1-2\beta)$ and 
\begin{equation}
A_3={2\beta(\gamma-1)\over (\gamma-2\beta)}\left(2\nu(\gamma-2\beta)\over
2\gamma\beta(1-2\beta)\right)^{\gamma\over 2\beta}
\label{athreefitted}
\end{equation} 

Using this mechanism, in the third column of Table \ref{U5}, 
we show the results for the three-dimensional Ising model.
The values of the critical exponents that we have used are
the best values reported in the literature\cite{Pelisseto}. i.e. 
$\gamma= 1.2372$, $\beta=0.3265$ and $\nu=0.6301$. 


\begin{table}[h]
\caption{\label{U5}\small Numerical values of expansion coefficients for the 
three-dimensional Ising universality class: $HT$ results taken from 
reference\cite{Pelisseto}, one loop($RG1$) and fitted exponents ($RGA$) results. 
The resulting values for the coefficients $r_{2n}$ were obtained from the relation 
(\ref{fversusr}) and using the values of $f_n^{\infty}$.}
\begin{ruledtabular}
\begin{tabular}{l|l|l|l}
             & $HT$       & $RG1$   & $RGA$ \\
\hline\hline
$f_0^\infty$ & 0.6024(15)    &   0.606     & 0.596   \\  
$f_1^\infty$ &      &   0.696         & 0.793   \\
$f_2^\infty$ &      &   0.6         & 0.613   \\
$f_3^\infty$ &     &   0.223         & 0.151  \\
$f_4^\infty$ &     & -0.066         & -0.084  \\
$f_5^\infty$ &     &   -0.0454  & 0.00527 \\  
$r_6$        & 2.056(5)      &  2.5               & 1.938 \\
$r_8$        & 2.3(1)   &  5.833    & 2.505  \\  
$r_{10}$     & -13(4)    &   -17.5  & -12.599 \\
$r_{12}$     &            & -192.5  & 10.902 \\
$f_1^0$      & 1.0527(7)           &    1.034             & 1.05  \\  
$f_2^0$      & 0.0446(4)           &    0.029             & 0.043 \\
$f_3^0$      & -0.0254(7)           &    -0.0034             & -0.0054  \\
$f_4^0$      &            &    0.0007             & 0.0013  \\
$f_5^0$      &            &    -0.00019             & -0.0004 \\
$f_1^{\rm c} $ &  0.9357(11)        &     0.96            & 0.939   \\
$f_2^{\rm c} $ &  0.08(7)        &     0.048            & 0.076   \\
$f_3^{\rm c} $ &          &    -0.0112             & -0.023 \\
$f_4^{\rm c} $ &          &    0.0049             &  0.0138 \\
$f_5^{\rm c} $ &          &    -0.0028             &  -0.011     \\
$R_4^+ $       &  7.81(2)        &     6.892            &  7.981 \\
$F_0^\infty$   &  0.03382(15)        &     0.0347            &  0.0263  \\

\end{tabular}
\end{ruledtabular}
\end{table}

Most of the values found with fitted exponents substantially improve
the value compared with the one-loop approximation in those cases
where a comparison can be made with HT expansions. This is
particularly notable in $r_6$, $r_8$ and $r_{10}$. $f_0^\infty$ and
$F_0^\infty$ are a little puzzling in this respect. However, it is
worth noting the experimental result reported in 
reference\cite{Pelisseto} where
$f^\infty_0=0.5917$, this being substantially different to the theoretical
value of $f^\infty_0=0.6369$\cite{Pelisseto}. The only other notable change is
that of $f_5^\infty$, which is related to $r_{12}$, where there is a
sign change passing from the one-loop to the adjusted values.

\begin{figure}[h]
\begin{center}
\includegraphics[width=3.5in,angle=0]{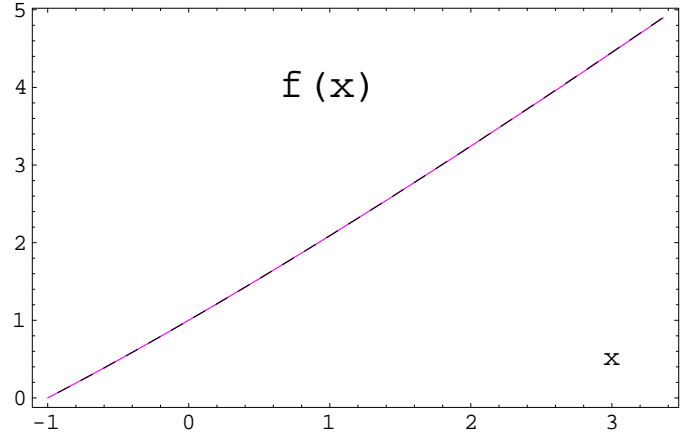}
\end{center}
\caption{\small The scaling function $f(x)$ of the three-dimensional 
Ising model using fitted exponents. The dashed line was taken from reference 
\cite{Guida}. Differences are not visible at this scale.}   
\label{fscalinggraph2}
\end{figure}

\begin{table}[h]
\caption{\label{U6}\small Numerical values of expansion coefficients
for the two-dimensional Ising universality class; high precision ($HT$) results taken from
reference\cite{Pelisseto} and results from using fitted exponents($RGA$)}.
\begin{ruledtabular}
  \begin{tabular}{l|l|l}
             & $HT$       & $RGA$ \\
\hline\hline
$f_0^\infty$ &  0.14753  &  0.1043   \\  
$f_1^\infty$ &   & 0.1696 \\
$f_2^\infty$ &   & 0.2365 \\
$f_3^\infty$ &   & 0.2931  \\
$r_6$        &  3.67867(7)          &  2.8571   \\
$r_8$        &  26.041(11)          &  15.2381  \\  
$r_{10}$     &  284.5(2.4)               &  125.714  \\
$r_{12}$     &  $4.2(7)\times 10^3$      &  1436.73  \\
$f_1^0$ &   & 1.1724 \\
$f_2^0$ &   & 0.1473 \\
$f_3^0$ &   & -0.0164 \\
$f_1^c$ &   & 0.785 \\
$f_2^c$ &   & 0.295 \\
$R_4^+ $       &  7.336774(10)      &    9.7594     \\ 
$F_0^\infty$   &  $5.92357(6)\times 10^{-5}$    &     0.10067     \\
\end{tabular}
\end{ruledtabular}
\end{table}




In the same way, we can also substitute the well-known exact values for the critical
exponents in the two-dimensional case, i.e. $\gamma=7/4$, $\beta=1/8$ and $\nu=1$.
Table \ref{U6} shows the subsequent results. In this case, except for the value 
of $f_0^\infty$, the values found for $r_{2n}$ show significant differences. This probably
hints at the inadequacy of the ansatz for the crossover functions (\ref{fittedwilson}).
We have not been able to find reported values for the coefficients $f_n^0$ and $f_n^c$, 
some of which are reported in Table \ref{U6}. It would be interesting to be able
to make the comparison.

\section{Conclusions}

Using environmentally friendly renormalization we derived a formal
expression for the equation of state for the $O(N)$ model. This
expression has the advantages that: i) it is \emph{derived} from an
underlying microscopic (field-theoretic) model; ii) requires as input
for any calculation only the three crossover Wilson functions
associated with a magnetization-dependent renormalization of the
field, $\varphi$, the composite operator, $\varphi^2$, and the coupling
constant, $\lambda$. In particular \emph{no} experimental input is
required on order to fix parameters; iii) it is parameterized 
by two non-linear scaling
fields - the transverse mass, and an anisotropy parameter closely
related to the stiffness constant; iv) it manifestly expresses all
relevant, desired analyticity properties, both in the critical region
and on the coexistence curve; v) universal coefficients associated
with the expansion of the equation of state in \emph{any} of the
asymptotic regimes may be simply calculated thereby obtaining
coefficients that are presently unknown.

After deriving one-loop expressions for general $N$, the formulation
was then used to calculate a parameterized, analytic expression for the
equation of state for $N=1$ for $2<d<4$. For $d=3$ it was shown how
the parameterization could be dispensed with and a closed-form,
non-parameterized expression for the equation of state derived.
Various universal coefficients associated with the asymptotic regimes
$x=0$, $x=\infty$ and $x=-1$ were derived for arbitrary $d$. For $d=3$
these were compared with previous results.

By taking the functional form for the Wilson functions at one-loop and
introducing constants to fit to the best-known asymptotic values of
the critical exponents, comparison was made for $d=3$ and $d=2$
between our calculated expansion coefficients and those, where known,
as derived using HT expansions, Monte Carlo, etc., in the different
asymptotic regimes. In most cases the fitted values were substantially
better than the one-loop values, agreement being better for $d=3$ than
$d=2$.

When continued to complex values of the external magnetic field the
universal equation of state should also capture the Lee-Yang edge.
There are additional points of non-analyticity in our expressions at
$z^2_{LY}=-2/(d-2)$ or $z^2_{LY}=-\nu/\beta$. These imaginary values of
$z$ are naturally associated with the Lee-Yang edge singularities. 
However, our equation of state is not yet optimized to include the 
associated crossover and we do not expect to obtain good estimates
for the associated universal exponents or amplitudes. Our formulation
can be adjusted to include this additional singularity, but as of yet
we have not studied the effect of including this crossover. We hope to
return to this in the future.

\vskip 0.3truein

\section*{Acknowledgements:} 
\vglue -0.2truein
CRS would like to thank DIAS and DGAPA, 
UNAM for financial support during part of this research. 
JAS would like to thank Denjoe O'Connor for hospitality at DIAS where part of
this work was done and PROMEP Mexico for financial support.

\end{document}